# Invasive advance of an advantageous mutation: nucleation theory

Lauren O'Malley[a], James Basham[a], Joseph A. Yasi[a,c], G. Korniss[a]
Andrew Allstadt[b] and Thomas Caraco[b,*]

[a] *Department of Physics, Applied Physics, and Astronomy, Rensselaer Polytechnic Institute, 110 8th Street, Troy, NY 12180-3590, USA*
[b] *Department of Biological Sciences, University at Albany, Albany, NY 12222, USA*

**Abstract**

For most organisms with viscous population structure, spatially localized growth drives the invasive advance of a favorable mutation. We model a two-allele competition where recurrent mutation introduces a genotype with a rate of local propagation exceeding the resident's rate. We capture ecologically important properties of the rare invader's stochastic dynamics by assuming discrete individuals and neighborhood interactions. To understand how individual-level processes may govern population patterns, we invoke the physical theory for nucleation of spatial systems. Nucleation theory discriminates between single-cluster and multi-cluster dynamics. A sufficiently low mutation rate, or a sufficiently small environment, generates single-cluster dynamics, an inherently stochastic process; a favorable mutation advances only if the invader cluster reaches a critical radius. For this mode of invasion we identify the probability distribution of waiting times until the favored allele advances to competitive dominance, and we ask how the critical cluster size varies as propagation or mortality rates vary. Increasing the mutation rate or system size generates multi-cluster invasion, where spatial averaging produces nearly deterministic global dynamics. For this process, an analytical approximation from nucleation theory, called Avrami's Law, describes the time-dependent behavior of the genotype densities with remarkable accuracy.

[*] Corresponding author.
[c] Present address: *Department of Physics, University of Illinois, 1110 West Green Street, Urbana, IL 61801-3080*
E-mail addresses: omalll@rpi.edu (L. O'Malley), bashaj@rpi.edu (J. Basham), yasi@uiuc.edu (J.A. Yasi), korniss@rpi.edu (G. Korniss), aa6545@albany.edu (A. Allstadt), caraco@albany.edu (T. Caraco)

# 1. Introduction

Understanding how localized biotic interactions govern invasion dynamics and an invader's subsequent spatial expansion remains a fundamental challenge in population biology (Levin et al., 1997; Wilson, 1998). Fisher (1937) and Kolmogorov et al. (1937) provided initial insight; they demonstrated how spatially structured dispersal can organize the advance of a favorable mutation. Fisher (1937) approximated the genetic process with a reaction-diffusion equation, and diffusion theory has since served repeatedly to model spatial expansion in ecology, evolution and epidemiology (see reviews by Fife, 1979; Okubo, 1980; Holmes et al., 1994; Murray, 2003). Generalizations beyond basic reaction-diffusion theory, intended to enhance biological realism, include models with age-dependent birth and death rates (Frantzen and van den Bosch, 2000; Neubert and Caswell, 2000), non-normal dispersal kernels (Kot at al., 1996; see Chesson and Lee, 2005), spatial heterogeneity (Cantrell and Cosner, 1991), and granularity of space or time (Neubert et al., 1995). Most of these analyses neglect demographic stochasticity, which can be important at local introduction of a rare type, to focus on the velocity of a travelling wave propelling a successful invader's advance (van den Bosch et al., 1990; Caraco et al., 2002; see Lewis and Pacala, 2000). Metz et al. (2000) present a useful review and guide to biological generalizations of diffusion processes.

Diffusion models sometimes predict velocities of spatial advance with accuracy, from the scale of nearest-neighbor infection (Zadoks, 2000) to biogeographic range expansion (van den Bosch et al., 1992). But travelling waves permit infinitely small population densities (van Baalen and Rand, 1998); consequently, the model may fail to capture essential properties of the dynamics of rarity (Durrett and Levin, 1994a), both at introduction and at the edge of an invader's expansion (Ellner et al., 1998; Thomson and Ellner, 2003). An alternative perspective on invasion assumes discrete individuals; individual-based models can directly address the impact of spatially clustered growth on the fate of a rare mutation when births and deaths occur as a random process (Claessen and de Roos, 1995; Iwasa et al., 1998; O'Keefe and Antonovics, 2002; Wei and Krone, 2005; Caraco et al., in press).

We analyze the spatial dynamics of a two-allele system when discrete individuals compete preemptively. We assume two-way, recurrent mutation in a population propagating clonally; the individual-level assumptions are simple, but the system's behavior can be complex. Mutation, introducing a superior allele, is a rare process, but can potentially occur anywhere in the spatial system. Each introduction initiates a separate cluster of invaders; a cluster may disappear through mortality, or may grow large. Our main results show how patterns in an advantageous mutation's spatial clustering influence the time elapsing until the superior allele replaces the resident. In particular, we present a novel dependence of an inferior allele's "lifetime" on the probabilistic rate of mutation.

To address the system's population-level behavior, we apply the theory for homogeneous nucleation of spatial systems (Kolmogorov, 1937; Johnson and Mehl, 1939; Avrami, 1940). Nucleation theory offers novel characterizations of the genetic or ecological clustering found in viscous populations (Gandhi et al., 1999; Korniss and Caraco, 2005). Foremost, we want to predict the time-dependent, global genotype densities, which are driven by locally clustered growth of the favored allele. To do so, we employ a powerful approximation, called Avrami's law (Ishibashi and Takagi, 1971; Rikvold et al., 1994; Korniss et al., 1999; Ramos et al., 1999) that describes our simulated mutation-selection dynamics quite accurately. We offer some new predictions about temporal behavior of spatially structured competitive systems, and suggest a framework for developing insights about spatially explicit invasion dynamics.

We list our model's assumptions, and then summarize a mean-field analysis and a pair approximation to the spatial model. Details of the latter, and some results concerning approach to extinction in our spatial model, are placed in appendices. Next we contrast single-cluster and multi-cluster invasion processes, motivating the application of nucleation theory. In a separate section, we use the theory to interpret simulation results. Finally, we suggest further application of nucleation theory in evolutionary ecology.

## 2. Spatial model

We consider an $L \times L$ lattice with periodic boundary conditions; a lattice site represents the minimal level of local resources required to sustain a single organism. Hence each site is either empty or occupied by one haploid individual (a resident or an invader). The local occupation numbers at site $\mathbf{x}$, $n_g(\mathbf{x}) = 0, 1$; $g = 1, 2$, count the number of resident and invader genotypes, respectively.

Competition for space is preemptive (Amarasekare, 2003; Shurin et al., 2004; Tainaka et al., 2004); an individual of either genotype may propagate clonally only if one or more of the $\delta$ nearest neighboring sites is empty. Specifically, if site $\mathbf{x}$ is empty, that site is colonized by allele $g$ at total probabilistic rate $\alpha_g \eta_g(\mathbf{x})$. $\alpha_g$ is the per-individual propagation rate of allele $g$. $\eta_g(\mathbf{x})$ is the density of allele $g$ on $\sigma(\mathbf{x})$, the set of nearest neighbors of site $\mathbf{x}$. Since the size of the interaction neighborhood is $|\sigma(\mathbf{x})| = \delta$, we have:

$$\eta_g(\mathbf{x}) = (1/\delta) \sum_{\mathbf{x}' \in \sigma(\mathbf{x})} n_g(\mathbf{x}'). \tag{1}$$

Most of our analyses below take $\delta = 4$, the four nearest neighbors on a square lattice. Unless noted explicitly in the figures, results are for $\delta = 4$. To test the robustness of some of our findings, we also considered neighborhood sizes $\delta = 8$ and $12$.

When the invader allele has a reproductive advantage, $\alpha_2 > \alpha_1$; we shall assume this condition holds when we apply nucleation theory to the dynamics (sections 4, 5 and 6). We assume two-way, recurrent mutation. Each individual of genotype $g$ independently mutates to genotype $(1 + |g - 2|)$ at constant probabilistic rate $\phi_g$. Finally, each individual carrying allele $g$ suffers mortality at probabilistic rate $\mu_g$.

We have a particular interest in the way clonal propagation drives local competition and cluster-size dynamics (O'Malley et al., 2005). Therefore, we let $\mu_1 = \mu_2 = \mu$, and $\phi_1 = \phi_2 = \phi$. The sole difference between genotypes is the difference in propagation rates. We summarize the local transition rates for an arbitrary site $\mathbf{x}$ as:

$$0 \xrightarrow{\alpha_1 \eta_1(\mathbf{x})} 1, \quad 0 \xrightarrow{\alpha_2 \eta_2(\mathbf{x})} 2, \quad 1 \xrightarrow{\mu} 0, \quad 2 \xrightarrow{\mu} 0, \quad 1 \xleftrightarrow{\phi} 2; \tag{2}$$

where 0, 1, 2 indicates whether the site is empty, occupied by the resident genotype, or occupied by an invader genotype, respectively. We list the paper's symbols in Table 1.

## 3. Mean-field model, pair approximation, and equilibrium phase diagram

We establish a context for our application of nucleation theory by summarizing a mean-field (MF) analysis and a pair approximation (PA) to the detailed model, together with the results of Monte Carlo simulations (see implementation details in the next section) indicating the equilibrium phase diagram of the model. The MF model assumes homogeneous mixing, and offers useful comparison to the behavior of the spatially structured process (Duryea et al., 1999; Neuhauser and Pacala, 1999; Bolker et al., 2000). The time-dependent global densities of the respective alleles are:

$$\rho_g(t) = (1/L^2) \sum_{\mathbf{x}} n_g(\mathbf{x}, t); \quad g = 1, 2. \tag{3}$$

Given the local transition rates in Expression (2), the MF equations are:

$$d\rho_1/dt = \alpha_1 \rho_1 (1 - \rho_1 - \rho_2) - \mu \rho_1 + \phi(\rho_2 - \rho_1),$$

$$d\rho_2/dt = \alpha_2 \rho_2 (1 - \rho_1 - \rho_2) - \mu \rho_2 + \phi(\rho_1 - \rho_2). \tag{4}$$

To highlight competition between the two alleles, we temporarily ignore mutation and set $\phi = 0$. Equations (4) exhibit three equilibrium points $(\rho_1^*, \rho_2^*)$. None permits coexistence; pre-emptive competition treats the genotypes as having completely overlapping niches (Amarasekare, 2003). For $\alpha_1 < \mu$ and $\alpha_2 < \mu$, neither of the two alleles can persist dynamically. For this case, mutual extinction is the stable stationary point, and $(\rho_1^*, \rho_2^*) = (0, 0)$. The critical propagation rate $\alpha_c$, which must be exceeded for monomorphic persistence under the MF assumption, is $\alpha_c = \mu$. We can anticipate that spatially clustered growth will increase the requirement for persistence, so that $\alpha_c > \mu$ (Neuhauser, 1998).

Keeping $\mu$ fixed, for $\alpha_1 > \mu$ and $\alpha_1 > \alpha_2$, the "resident" competitively excludes the "invader" at equilibrium. In this case $(\rho_1^*, \rho_2^*) = (1 - \mu/\alpha_1, 0)$ is stable. If $\alpha_2 > \mu$ and $\alpha_1 < \alpha_2$, the invader allele prevails and excludes the resident, and $(\rho_1^*, \rho_2^*) = (0, 1 - \mu/\alpha_2)$ is stable. Consequently, the MF model predicts that when $\alpha_2 > \alpha_1 > \mu$, the resident allele, resting at its positive equilibrium density, will be invaded by the alternate allele. The invader will then advance to exclude the resident competitively (to order $\phi$ in the presence of mutation).

Pair approximation (Matsuda et al., 1987; Rand, 1999; Satō and Iwasa, 2000; van Baalen, 2000) incorporates correlations of nearest neighbors into the dynamics, and consequently may predict equilibrium densities more accurately than MF models (Ellner et al., 1998; Caraco et al., 2001; Tainaka, 2003). Appendix A presents a pair approximation to our model. In the invader's absence (with $\phi = 0$), PA predicts an equilibrium global density for the resident allele that depends on neighborhood size, $\rho_1^* = [\delta - 1 - \delta(\mu/\alpha_1)]/[\delta - 1 - (\mu/\alpha_1)]$; simulations indicate improvement over the mean-field prediction. Compared to the MF result, PA (for fixed $\mu$) increases the critical value of the clonal propagation rate $\alpha_c$ in the phase transition between extinction and persistence. Under PA $\alpha_c = \mu\delta/(\delta - 1)$, so that $\alpha_c$ decreases as neighborhood size increases. Spatial clustering increases the probability that a site neighboring an individual is occupied beyond the occupation probability of a randomly chosen site. Consequently, persistence under PA requires a greater propagation rate than is required under homogeneous mixing, since propagules are wasted on occupied neighboring sites (Durrett and Levin, 1994a). But PA preserves the MF model's linear behavior in the vicinity of the critical point, $\rho_g^* \approx (\alpha_g - \alpha_c)/\mu$ (see Appendix B).

To explore the spatial model's equilibrium behavior, we carried out Monte Carlo (MC) simulations on $L \times L$ square lattices with periodic boundaries. The equilibrium phase diagram, Fig. 1(A), agrees qualitatively with results of the MF and PA. The two positive equilibria (where one allele excludes its competitor) exhibit a transition (through a very narrow, continuous crossover) across the line $\alpha_1 = \alpha_2$, reflecting symmetry of the local dynamics; see Fig. 1(B). The fully spatial system also exhibits mutual extinction for $\alpha_1 < \alpha_c$ and $\alpha_2 < \alpha_c$ (where $\alpha_c$ depends on $\mu$). The transition between positive density and extinction (alleles cannot advance when $\alpha_g < \alpha_c$) is continuous. The behavior of the resident's equilibrium global density near the critical point, $\alpha_c \approx 1.65\mu$ (Grassberger and de la Torre, 1979), is described by a non-trivial power-law. That is, $\rho_1^* \propto (\alpha_1 - \alpha_c)^\beta$ with $\beta \approx 0.58$; see Fig. 1(C) (by symmetry, the invader shows the same behavior). Hence neither the MF nor PA model captures quantitative detail of the dynamics near extinction; both underestimate the propagation rate where extinction occurs, and just as importantly, fail to predict the correct power-law near the transition. Appendix B explores the nature of the transitions between the three equilibrium phases, and includes an expanded comment on competitive exclusion in the presence of mutation. Appendix B also clarifies the relationship between our spatial model's extinction dynamics and the contact process (Durrett and Levin, 1994b), indicating the general ecological significance of dispersal-limited extinction (Oborny et al., 2005).

## 4. Critical cluster size, single-cluster and multi-cluster invasion

Our focal analyses concern the time-dependent behavior of the two-allele system far away from dispersal-limited extinction. Each allele can persist dynamically in the other's absence, but allele 2, the invader, has a clonal-propagation advantage. We assume that mutation is rare, and restrict attention to the parameter regime where $\phi \ll \mu < \alpha_1 < \alpha_2$ and $\phi \ll (\alpha_2 - \alpha_1)$, i.e., the region where the invader will eventually exclude the resident competitively.

We performed extensive simulations of our model's invasion dynamics. The time unit was one Monte Carlo step per site (MCSS) during which $L^2$ lattice sites were chosen randomly and updated probabilistically. These procedures mimic continuous-time stochastic dynamics in the large $L$ limit (Durrett and Levin, 1994b; Korniss et al., 1999). We initialized each simulation with every site occupied by the resident, allele 1. The invader (allele 2) entered through rare, recurrent mutation.

Nucleation theory predicts that the invader allele will statistically tend to advance only if it generates a cluster sufficiently large that the probability it will grow at its periphery exceeds ½. For realistic spatial scales, there should exist a critical radius for clusters of the invader; smaller clusters tend to decrease in size, and larger clusters tend to grow (Gandhi et al., 1999; Korniss et al., 1999; O'Malley et al., 2005). Snapshots of simulation configurations reveal strongly clustered invader growth, and confirm the existence of a critical cluster size $R_c$, which we estimate below.

For given parameter values, there also exists a characteristic length scale $R_d$, the typical distance separating invading clusters. For $L \ll R_d$, invasion almost always occurs through the spread of a single invading cluster (single-cluster invasion), while for $L \gg R_d$, the invasion results from many invading clusters (multi-cluster invasion). Conversely, when we fix linear system size $L$ and all other parameters except the mutation rate $\phi$, the characteristic length scale $R_d$ becomes a function of $\phi$. For this case, there is a characteristic value of $\phi$ such that for a sufficiently rare occurrence of mutation, multi-cluster invasion by the favored allele crosses over to the single-cluster geometry. These two different invasion modes, single-cluster and multi-cluster, are illustrated by snapshots of configurations in Figs. 2 and 3, respectively. Importantly, these different modes of spatial growth are associated with drastically different sets of dynamical properties, as can be seen from the time series of the global densities in Fig. 4. In the single-cluster regime, the waiting time until the first (and almost always the only) successful invader cluster appears is inherently stochastic [Fig. 4(A)]; the variance of the time elapsing until the invader outnumbers the resident (the system lifetime) is comparable to its mean. In contrast, in the multi-cluster regime, global densities exhibit a near-deterministic lifetime [Fig. 4(B)].

For either invasion mode, we consider the invader allele competitively dominant when the resident allele's global density first falls to one-half its quasi-equilibrium density (in the absence of the invader) $\rho_1^*$. That is, the invader dominates when $\rho_1(t)|_{t=\tau} = \rho_1^*/2$. Given that we gauge the invader allele's advance by quantifying the resident's decay, we define the non-negative random variable $\tau$ as the first-passage time of the resident's density to $\rho_1^*/2$. We refer to the expected value $\langle\tau\rangle$ as the resident allele's "metastable" lifetime (Rikvold et al., 1994; Korniss and Caraco, 2005).

## 5. Homogeneous nucleation and growth

The preceding observations justify applying the theory of homogeneous nucleation and growth (Kolmogorov, 1937; Johnson and Mehl, 1939; Avrami, 1940) to predict important characteristics of the spread of an advantageous allele. This framework, commonly referred to as the KJMA theory or Avrami's Law, has successfully described conceptually analogous dynamic phenomena in physics (Ishibashi and Takagi, 1971; Duiker and Beale, 1990; Rikvold et al., 1994; Richards et al., 1995; Ramos et al., 1999; Korniss et al., 1999), chemistry (Ben-Naim and Krapivsky, 1996; Karttunen et al., 1998; Machado et al., 2005), DNA replication (Herrick et al., 2002), and certain spatially structured ecological interactions (Korniss and Caraco, 2005).

A simple application of KJMA theory to our model assumes that nucleation of a successful invading cluster is a Poisson process with a constant (in both space and time) nucleation rate per unit area $I$. Further, the theory assumes that once a successful cluster of the advantageous allele has been nucleated, it grows deterministically with a constant radial velocity $v$. Here, we briefly present the main features of Avami's Law and then summarize a scaling argument (Rikvold et al., 1994; Ramos et al., 1995) that identifies the time and length scales governing single-cluster versus multi-cluster invasion. For a detailed derivation of KJMA theory, the reader may consult, e.g., Ishibashi and Takagi (1971), Korniss and Caraco (2005), or the original papers cited above.

For small spatial systems, where $L \ll R_d$ ($R_d$ to be determined), the invasive allele advances as a single invading cluster, the first cluster that nucleates and sweeps through the environment before another appears (Rikvold et al., 1994; Richards et al., 1995; Korniss and Caraco, 2005). Since nucleation of a successful invasive cluster is a Poisson process, the cumulative probability distribution of invasion times $P_{not}(t) = \Pr[\tau > t]$, i.e., the probability that the resident allele's global density has not decayed to $\rho_1^*/2$ by time $t$ (governed by the first and only invading cluster in the single-cluster regime), is exponential:

$$P_{not}(t) = \begin{cases} 1 & \text{for } t \leq t_g \\ \exp[-(t-t_g)/\langle t_n \rangle] & \text{for } t > t_g \end{cases}. \tag{5}$$

Here, $\langle t_n \rangle = (L^2 I)^{-1}$ is the average time elapsing until nucleation occurs, and $t_g \sim L/v$ is the approximately deterministic growth time until the invading allele drives the resident to half its initial density. For very small nucleation rates per unit area, we have $\langle t_n \rangle \gg t_g$. Therefore, the lifetime of the resident allele is governed by the large average waiting times until the first successful cluster of the invasive allele nucleates, so that $\langle\tau\rangle = \langle t_n \rangle + t_g \approx \langle t_n \rangle$.

For large spatial systems, $L \gg R_d$, so that system size exceeds the typical distance between invader clusters. Therefore, many randomly nucleated and expanding clusters of the favored allele contribute to the resident allele's decline. In the limit as $L \to \infty$, the global dynamics approaches deterministic functions. The number of invader clusters grows large, so that spatial averaging of their behavior strongly limits stochastic fluctuations at the global level. As a direct consequence, $P_{not}(t)$ approaches a step-function centered on the system-size independent lifetime $\langle\tau\rangle$. In this large-system limit, $R_c \ll R_d \ll L$, and the global densities can be approximated closely by Avrami's Law applied to a two-dimensional system (Kolmogorov, 1937; Johnson and Mehl, 1939; Avrami, 1940):

$$\rho_1(t) \approx \rho_1^* \exp\left[-\ln(2)\left(\frac{t}{\langle\tau\rangle}\right)^3\right]$$

$$\rho_2(t) \approx \rho_2^*\left(1 - \exp\left[-\ln(2)\left(\frac{t}{\langle\tau\rangle}\right)^3\right]\right) \quad (6)$$

One can extract the scaling behavior of the resident allele's system-size independent lifetime $\tau \equiv \langle\tau\rangle$ under multi-cluster invasion (Rikvold et al., 1994; Ramos et al., 1999). The average diameter of invader clusters at time $t = \tau$ scales as $v\tau$, assuming a constant radial growth velocity. Considering the size to which a cluster can grow without contacting another, the average spatial separation between the centers of invader clusters should scale similarly at time $t = \tau$, thus $R_d \sim v\tau$. Further, at the same instant, on average, one cluster has nucleated in a corresponding area of diameter $R_d$; thus, in two dimensions, $IR_d^2\tau \sim 1$. Hence, one obtains:

$$\tau \sim (Iv^2)^{-1/3}, \quad (7)$$

and

$$R_d \sim \left(\frac{v}{I}\right)^{1/3}. \quad (8)$$

Expression (7) gives the characteristic time and expression (8) gives the characteristic length scale of the system in the limit as $L \to \infty$. We infer the following from expression (7). The parameters of a specific model for spatially structured ecological interactions (here, local transition rates $\phi$, $\alpha_1$, $\alpha_2$, $\mu$, and $\delta$) govern the characteristic time scale (the lifetime $\tau$) through their impact on $I$, the superior allele's nucleation rate per unit area of the superior allele, and on $v$, the radial growth velocity of the invading clusters. Expression (7) suggests that the time elapsing until the invader dominates the resident will vary with the inverse of the cube root of the rate at which invader clusters reach critical size (i.e., nucleate); for discussion, see Korniss and Caraco (2005).

## 6. Simulation results and interpretation

We conducted separate simulations to ask how model parameters affected critical cluster size, hence the critical radius $R_c$. Recall that at critical size, growth and decay of a cluster have equal probability. For this analysis only, we began each simulation by placing a single favored-allele cluster, of a selected size [and with a "round" shape, $A(0) \sim R^2(0)$], on the lattice. We set the mutation rate $\phi = 0$, so the invader cluster's dynamics depended on propagation and mortality only. If the invader declined to extinction, we recorded the elapsed time. Otherwise, the invader advanced to competitive dominance, and we recorded the time at which the resident's global density first decayed to $\rho_1^*/2$ (invasion time).

For each initial cluster size and ($\alpha_g$, $\mu$)-combination, we plotted the fraction of $10^4$ simulations where the favored allele invaded the resident; by definition, this fraction (the relative success rate) is ½ at the critical cluster size. For mortality rates $\mu = 0.1$ and $\mu = 0.2$, Fig. 5(A) shows how the relative success rate changes as the resident's propagation rate $\alpha_1$ increases toward the invader's propagation rate $\alpha_2$, and implies, in turn, how the critical cluster size increases at the same time. Fig. 5(A) also indicates that doubling the mortality rate ($\mu = 0.1$ to $\mu = 0.2$) had little influence on critical cluster size for the given $\alpha_g$. However, greater mortality reduced the resident allele's metastable lifetime. Fig. 5(B) shows that when the favored allele advanced, greater mortality nearly halved the mean time for the resident's decay; the global dynamics was accelerated.

We anticipated that $R_c$ would increase as the difference between propagation rates ($\alpha_2 - \alpha_1$) declined. The increased mortality rate opened more sites for neighborhood growth of the superior allele, and so relaxed the constraints of preemptive competition (Korniss and Caraco, 2005).

*Single-cluster invasion*

To analyze single-cluster invasion, we fixed $\alpha_1 = 0.5$, and $\alpha_2 = 0.7$, for various system sizes $L = 16, 32, 64$. For three values of the mortality rate, $\mu = 0.1, 0.15$, and $0.2$, we consider mutation rates $10^{-7} \leq \phi \leq 10^{-5}$. Single-cluster invasion is inherently stochastic, and must be analyzed probabilistically. While local mutation occurs as a Poisson process, it is not known *a priori* if the nucleation of a "supercritical" cluster will also be Poisson. Therefore, for simulations in the single-cluster regime, we constructed cumulative probability distributions for the lifetime of the resident allele $P_{not}(t)$, i.e., the probability that the global density of the resident has not yet declined below $\rho_1^*/2$ by time $t$. We found that the distributions are indeed exponentials, in accordance with Eq. (5). We show results for a fixed (sufficiently small) system size and mortality rate for three mutation rates [Fig. 6(A)]. From the slopes of the exponential distributions we obtained the average nucleation times $\langle t_n \rangle$ [Fig. 6(B)], which, in turn, yields the mutation-rate dependence of the nucleation rate per unit area $I(\phi)$. Since $\langle t_n \rangle = (IL^2)^{-1}$, we conclude that for fixed linear system size $L$, $\langle t_n \rangle \sim \phi^{-1}$ and $I(\phi) \sim \langle t_n \rangle^{-1} \sim \phi$; the rate at which a critical cluster forms scales with the mutation rate. We also analyzed the system-size dependence (within the single-cluster regime) of the average nucleation time, again, by determining the slopes of the cumulative lifetime distributions for various system sizes ([Fig. 7(A)], indicating that the mean waiting time for nucleation of a critical cluster $\langle t_n \rangle \sim L^{-2}$ [Fig. 7(B)].

In the single-cluster regime, the advance of the invasive allele (and decline of the resident) exhibits significant stochastic variation; as noted above, the standard deviation of the resident's lifetime $\tau$ is as large as its mean. The spread of the advantageous mutation is initiated and completed by the first randomly nucleated, successful cluster of the favored allele. Any number of smaller clusters may fail to grow before the successful nucleation event occurs. For very low values of the mutation rate $\phi$, the lifetime is dominated by the very large expected waiting time for nucleation, hence:

$$\langle \tau \rangle = \langle t_n \rangle + t_g \approx \langle t_n \rangle \sim \phi^{-1}. \tag{9}$$

*Multi-cluster invasion*

In the multi-cluster regime the dynamics of the advantageous allele's global density behaves more predictably. Nucleation and growth of multiple clusters renders global densities sums of random variables; spatial averaging within the multi-cluster process reduces variability of the time-dependent global densities among realizations of the dynamics (Rikvold et al., 1994; Ramos et al., 1999; see Wilson et al., 1993). This "self-averaging" induces the nearly deterministic behavior of the global densities in the limit as $L \to \infty$. Therefore, we compared time-dependent global densities observed in our simulations to Avrami's Law, Eq. (6). Our results in Fig. 8 show that it is, indeed, an excellent approximation for times $t \leq \langle \tau \rangle$. At times significantly larger than $\langle \tau \rangle$ invading clusters begin to coalesce and percolation effects become important. Assuming that spreading velocity of the invading clusters is constant, KJMA theory, Eq. (7), predicts that the resident allele's metastable lifetime scales according to:

$$\langle \tau \rangle \sim [I(\varphi)]^{-1/3} \sim \phi^{-1/3}, \tag{10}$$

in the multi-cluster regime; note the important difference between proportionalities for single and multi-cluster invasion, cf. Eq. (9). The exponent estimated from our simulations, approximately $-0.3$, is close to the predicted value; see Fig. 9. Hence Avrami's Law described the time course of resident's metastable decay with reasonable accuracy. The deviation from prediction likely indicates that, contrary to our assumption, cluster radii, at least for early times, exhibit a deviation from linear growth before asymptotically reaching a constant radial spreading velocity, which may also be affected by nontrivial properties of the cluster perimeter (cf. Lewis and Kareiva, 1993).

To study robustness of the KJMA-type nucleation and spread of the mutant allele, we also considered neighborhood sizes beyond nearest neighbors ($\delta = 8$ and 12). The results for these are also shown in Figs. 8 and 9(A). Observed global densities are qualitatively similar to those for nearest-neighbor propagation ($\delta = 4$). The shape of the time-dependent global densities is still well approximated by Avrami's Law, Eq. (6), up to times comparable to $\langle \tau \rangle$ (Fig. 8), when effects of cluster coalescence become important, and nucleation theory breaks down. Interestingly, the lifetime $\langle \tau \rangle$ exhibits the same scaling with the mutation rate $\langle \tau \rangle \sim \phi^{-0.3}$ [Fig. 9(a)]. These results indicate that neighborhood size, hence the opportunity for local dispersal, affects the lifetime through the nucleation rate per unit area $I$ and through $v$, the radial growth velocity of each nucleated invader cluster.

## 7. Discussion

We have suggested a new framework for analyzing the invasion and spread of a recurring favorable mutation when competition between genotypes follows locally structured, probabilistic rules. Our study is somewhat distinct among models of spatial competition in that our application of nucleation theory stresses the system's time-dependent characteristics, rather than only the asymptotic behavior. We focused on generic features of the resident allele's lifetime and the global densities, both at a phenomenological level. Specifically, we emphasized the form of the probability distribution of the resident allele's decay to competitive exclusion in the single-cluster regime, and then emphasized the functional time-dependence of the global allelic densities in the multi-cluster regime. We found that nucleation theory, Avrami's Law in particular, describes the simulated dynamics very well. We detailed quantitative effects of varying both the size of the environment and the mutation rate. Finally, we indicated that local, measurable rates of propagation and mortality, along with neighborhood size, can affect the scaling of Avrami's Law, hence can affect the global dynamics.

Fig. 10 summarizes our essential results. For infinitely large systems and sufficiently small mutation rates, the system exhibits multi-cluster invasive spread and is well approximated by Avrami's Law. The self-averaging lifetime of the resident allele scales as $\langle \tau \rangle \sim \phi^{-0.3}$ (with other parameters fixed). For larger mutation rates, small invasive clusters almost immediately coalesce, and nucleation theory breaks down. In this regime, MF approximation begins to work reasonably well, since the more numerous small clusters mix almost immediately. For finite systems, however, there is a sufficiently small mutation rate $\phi$, such that the typical separation between invasive clusters would become larger than the linear system size ($R_d \gg L$). Here, the invasion dynamics crosses over to the single-cluster mode. For example, if $L = 64$, crossover to the single-cluster regime occurs when $\phi \sim 10^{-5}$, while for $L = 128$ we noted crossover when $\phi \sim 10^{-6}$ (Fig. 10). In this regime the average lifetime (together with its variance) scales as $\langle \tau \rangle \sim \phi^{-1}$.

The primary feature of our model is that rare random mutations occur independently in space and time, and then initiate invasive clusters. Only some reach a critical size, and continue to grow at their periphery. Once the radius of an invading cluster reaches sufficient length, one can anticipate that the radial velocity approaches its asymptotic value; see Lewis and Kareiva (1993) for deterministic analysis of diffusive ecological invasion in two dimensions (where an "Allee effect" produces another sort of critical radius). Further study of the stochastic interface separating the region of allele 2 from the region of allele 1, and an invader cluster's approach to the asymptotic spreading velocity may suggest finer approximations than Avrami's Law for our two-allele model.

We have demonstrated the utility of nucleation theory in a model where the inferior local competitor lacks a compensatory advantage of lower mortality (Caraco et al., in press) or a larger dispersal profile (Durrett and Levin, 1998; Bolker and Pacala, 1999). Nucleation theory suggests a context for an increased understanding of more complex spatial ecologies. Ideally, an analysis of spatial invasion would reconcile results of individual-based simulations with predictions deduced from tractable mathematical models. For example, Wilson (1998) resolves differences between simulation of a spatially structured predator-prey interaction and an elaboration of a corresponding diffusion model. As indicated above, we anticipate that detailed analysis of the critical radius will yield mechanistic hypotheses linking local interaction and population processes. Finally, the framework for analyzing invasive growth suggested here may be extended to more general problems involving temporal variation or spatial heterogeneity in demographic rates (Chesson 2000; Korniss and Caraco 2005; Seabloom et al. 2005).

**Acknowledgements**
We gratefully acknowledge discussions with Z. Rácz, P.A. Rikvold, M. A. Novotny, T. Ala-Nissila, Z. Toroczkai, I.-N. Wang and G. Robinson. We also thank Z. Toroczkai for helping us implement a numerical integrator routine for the mean-field and pair approximations. This material is based upon work supported by the National Science Foundation under Grant No. 0342689. G. K. was also supported in part by NSF through Grant DMR-0113049 and by the Research Corporation through Grant No. RI0761.

## Appendix A. Pair approximation

Pair approximation tracks the deterministic dynamics of both global densities and local densities conditioned on the state of a neighboring site. The text defines global densities $\rho_g$. For local densities, suppose that site **x** has elementary state $j$. $q_{k/j}$ is the conditional probability that a randomly selected site on the neighborhood of site **x** has state $k$. By conditional probability:

$$q_{0/j} + q_{1/j} + q_{2/j} = 1 \quad (j = 0, 1, 2), \tag{A.1}$$

where $j$ indicates empty (0), resident-occupied (1) and invader-occupied (2) sites; see Expression (2) in the text. The dynamics of local densities involves global pair densities $\rho_{jk}$, the unconditional probability that site **x** has elementary state $j$, and a randomly chosen nearest neighbor of site **x** has state $k$. We assume spatial symmetry, $\rho_{jk} = \rho_{kj}$, so that:

$$q_{k/j} = (\rho_k / \rho_j) q_{j/k}. \tag{A.2}$$

Equality constraints and spatial symmetry imply that five equations define the pair approximation; we write equations for $\rho_1$, $\rho_2$, $q_{1/1}$, $q_{1/2}$, and $q_{2/2}$. We can specify the remaining 7 variables in terms of the 5 variables the dynamics tracks. To do so, we follow Iwasa et al. (1998).

Under pair approximation the invader allele's global density evolves according to:

$$d\rho_2/dt = \alpha_2 q_{0/2} \rho_2 + \phi(\rho_1 - \rho_2) - \mu\rho_2. \tag{A.3}$$

The respective terms represent local growth, mutation, and mortality. $q_{0/2}$ is the fraction of the sites neighboring a site occupied by the invader that, on average, is empty. Rewritten in terms of the 5 dynamic variables, Eq (A.3) becomes:

$$d\rho_2/dt = \alpha_2 \rho_2(1 - q_{1/2} - q_{2/2}) + \phi(\rho_1 - \rho_2) - \mu\rho_2. \tag{A.4}$$

To write the dynamics of a local density, we require the dynamics of the corresponding doublet, a global density. First, consider the global-pair density $\rho_{22}$, the frequency of invader-allele pairs among all neighboring-site pairs. $\rho_{22}$ has dynamics:

$$d\rho_{22}/dt = 2\phi(\rho_{12} - \rho_{22}) - 2\mu\rho_{22} + 2\rho_{02}\left[\frac{\alpha_2}{\delta} + (\delta-1)\frac{\alpha_2}{\delta} q_{2/02}\right]. \tag{A.5}$$

The first two terms represent net mutation and mortality, respectively. The third represents local propagation into the empty site of empty-invader and invader-empty pairs. An invader occupies one site in each such pair, colonizing the empty site at rate $\alpha_2/\delta$. Each of the remaining neighbors of the empty site may, or may not, be occupied by an invader. The conditional probability that an invader allele occupies a third site neighboring the open site of an empty-invader pair is $q_{2/02}$. Ordinary pair approximation assumes that $q_{2/02} \approx q_{2/0}$, and replaces the former with the latter, allowing closure of the system of equations (Matsuda et al., 1992; van Baalen, 2000). Replacing $q_{2/02}$ with $q_{2/0}$ in equation (A.5), rewriting in terms of the state variables and rearranging yield:

$$d\rho_{22}/dt = 2\rho_2[\phi(q_{1/2} - q_{2/2}) - \mu q_{2/2}]$$
$$+ 2\rho_2(1 - q_{1/2} - q_{2/2})\left[\frac{\alpha_2}{\delta}\left(1 + [\delta-1]\frac{\rho_2}{1-\rho_1-\rho_2}[1 - q_{1/2} - q_{2/2}]\right)\right]. \tag{A.6}$$

Since $q_{i/i} = \rho_{ii}/\rho_i$, we note (Satō and Iwasa, 2000):

$$dq_{2/2}/dt = (1/\rho_2) d\rho_{22}/dt - (1/\rho_2) q_{2/2} (d\rho_2/dt)$$

Then, using equations (A.5) and (A.6), we have the conditional-density dynamics $dq_{2/2}/dt$:

$$dq_{2/2}/dt = 2(1 - q_{1/2} - q_{2/2})\left[\frac{\alpha_2}{\delta}\left(1 + [\delta - 1]\frac{\rho_2}{1 - \rho_1 - \rho_2}[1 - q_{1/2} - q_{2/2}]\right)\right]$$
$$+ 2\phi(q_{1/2} - q_{2/2}) - q_{2/2}\left[\alpha_2(1 - q_{1/2} - q_{2/2}) + \frac{\phi(\rho_1 - \rho_2)}{\rho_2} + \mu\right]. \quad (A.7)$$

The resident's global density evolves according to:
$$d\rho_1/dt = \alpha_1 q_{0/1} \rho_1 + \phi(\rho_2 - \rho_1) - \mu\rho_1. \quad (A.8)$$
The terms refer to local propagation, net effect of mutation, and mortality. Rewriting in terms of the 5 dynamic variables yields:
$$d\rho_1/dt = \alpha_1 \rho_1\left(1 - q_{1/1} - \left(\frac{\rho_2}{\rho_1}\right)q_{1/2}\right) + \phi(\rho_2 - \rho_1) - \mu\rho_1. \quad (A.9)$$

To obtain $d\,q_{1/2}/dt$, we begin with the doublet $\rho_{12}$:
$$d\rho_{12}/dt = \phi(\rho_1 q_{1/1} + \rho_2 q_{2/2}) - 2\mu\rho_2 q_{1/2}$$
$$+ \frac{\delta - 1}{\delta}\frac{\rho_1\rho_2}{1 - \rho_1 - \rho_2}\left[1 - \frac{\rho_2}{\rho_1}q_{1/2} - q_{1/1}\right](1 - q_{1/2} - q_{2/2})(\alpha_2 + \alpha_1). \quad (A.10)$$

Since $q_{1/2} = \rho_{12}/\rho_2$,
$$dq_{1/2}/dt = (1/\rho_2)d\rho_{12}/dt - (1/\rho_2)q_{1/2}(d\rho_2/dt).$$
Using equations (A.4) and (A.10), we have the conditional-density dynamics:
$$dq_{1/2}/dt = \phi\left(\frac{\rho_1}{\rho_2}q_{1/1} + q_{2/2}\right) - q_{1/2}\mu$$
$$+ \frac{\delta - 1}{\delta}\frac{\rho_1}{1 - \rho_1 - \rho_2}\left[1 - \frac{\rho_2}{\rho_1}q_{1/2} - q_{1/1}\right](1 - q_{1/2} - q_{2/2})(\alpha_2 + \alpha_1) \quad (A.11)$$
$$- q_{1/2}\left[\alpha_2(1 - q_{1/2} - q_{2/2}) + \frac{\phi(\rho_1 - \rho_2)}{\rho_2}\right].$$

To obtain $d\,q_{1/1}/dt$, we begin with the doublet $\rho_{11}$:
$$d\rho_{11}/dt = 2\phi(\rho_2 q_{1/2} - \rho_1 q_{1/1}) - 2\mu\rho_1 q_{1/1}$$
$$+ 2\rho_1\left[1 - \frac{\rho_2}{\rho_1}q_{1/2} - q_{1/1}\right]\left[\frac{\alpha_1}{\delta}\left(1 + [\delta - 1]\frac{\rho_1}{1 - \rho_1 - \rho_2}\left(1 - \frac{\rho_2}{\rho_1}q_{1/2} - q_{1/1}\right)\right)\right]. \quad (A.12)$$

Recognizing that
$$dq_{1/1}/dt = (1/\rho_1)d\rho_{11}/dt - (1/\rho_1)q_{1/1}(d\rho_1/dt)$$
leads to the conditional-density dynamics:
$$dq_{1/1}/dt = 2\left[1 - \frac{\rho_2}{\rho_1}q_{1/2} - q_{1/1}\right]\left[\frac{\alpha_1}{\delta}\left(1 + [\delta - 1]\frac{\rho_1}{1 - \rho_1 - \rho_2}\left(1 - \frac{\rho_2}{\rho_1}q_{1/2} - q_{1/1}\right)\right)\right]$$
$$+ 2\phi\left(\frac{\rho_2}{\rho_1}q_{1/2} - q_{1/1}\right) - q_{1/1}\left[\alpha_1\left(1 - q_{1/1} - \frac{\rho_2}{\rho_1}q_{1/2}\right) + \frac{\phi(\rho_2 - \rho_1)}{\rho_1} + \mu\right]. \quad (A.13)$$

This equation completes the pair-approximation system.

*Resident at Equilibrium*
If $\phi = \rho_2 = 0$ in equation (A.9) and find:

$$q_{1/1}^* = 1 - \mu/\alpha_1. \tag{A.14}$$

Using this result in equation (A.13), with $\phi = 0$ and $\rho_2 = 0$, yields the resident's equilibrium global density, in the absence of the advantageous allele, under pair approximation:

$$\rho_1^* = \frac{\delta - 1 - \delta(\mu/\alpha_1)}{\delta - 1 - (\mu/\alpha_1)}. \tag{A.15}$$

which is smaller than the MF equilibrium density, $\rho_1^* = 1 - \mu/\alpha_1$.

*Deterministic Condition for Initial Spread of the Advantageous Allele*
Now we assume that mutation already has introduced the invader allele, and set the mutation rate $\phi = 0$ in both genotypes; the analysis then approximates the expected outcome for an initial cluster of the invader allele (van Baalen, 2000).

Any increase in the invader allele's density requires $d\rho_2/dt > 0$. Evaluated at $\phi = 0$, this condition implies:

$$1 - q_{1/2} - q_{2/2} > \left(\mu/\alpha_2\right). \tag{A.16}$$

That is, given that an invader occupies an arbitrary site, the conditional density of empty sites on the invader's neighborhood must exceed the invader's mortality to propagation-rate ratio for the advantageous allele to advance when rare. We evaluate the conditional densities in expression (A.16) to predict the invader allele's advance when rare.

Proceeding as in Iwasa et al. (1998), we solve the dynamics in the local environment of a rare invader allele, using the resident-only equilibria, $\rho_1^*$ and $q_{1/1}^*$ with $\rho_2 = 0$. That is, given the results in (A.14) and (A.15), we find equilibrium values for $q_{1/2}$ and $q_{2/2}$ when the invader is rare. Finally, we use these equilibria to evaluate expression (A.16), the condition for positive growth of the invader's global density. Although the invader is globally rare, local clustering implies that $q_{2/2}$ need not be close to zero.

Using Eqq (A.14) and (A.15) in equations (A.7) and (A.11) yields:

$$q_{1/2} = \frac{\left(\delta - 1 - \frac{\delta\mu}{\alpha_1}\right)(1 - q_{1/2} - q_{2/2})\left(\frac{\alpha_2 + \alpha_1}{\delta}\right)}{\alpha_2(1 - q_{1/2} - q_{2/2}) + \mu}, \tag{A.17}$$

and

$$q_{2/2} = \frac{2(1 - q_{1/2} - q_{2/2})\left(\alpha_2/\delta\right)}{\alpha_2(1 - q_{1/2} - q_{2/2}) + \mu}, \tag{A.18}$$

where (A.17) and (A.18) assume dynamic equilibrium.

$(1 - q_{1/2} - q_{2/2})$ is the conditional density of empty sites, given that a rare invader occupies the focal site. We know that the quantity $q_{0/2} = (1 - q_{1/2} - q_{2/2})$ must exceed $\mu/\alpha_2$ for the invader allele to advance when rare, by expression (A.16). Therefore, the sum $(\mu/\alpha_2 + q_{1/2} + q_{2/2})$ must be less than 1, if the rare allele is to advance. Paralleling Iwasa et al. (1998), the condition for advance of the invader allele when rare, from expression (A.16), is:

$$\frac{\mu}{\alpha_2} + \frac{1}{\delta} + \frac{1}{2\alpha_2}(\alpha_2 + \alpha_1)\left(\frac{\delta - 1}{\delta} - \frac{\mu}{\alpha_1}\right) < 1. \tag{A.19}$$

Expression (A.19) holds, and the rare allele invades, only if $\alpha_2 > \alpha_1$. By construction, PA yields the same deterministic criterion for invasion by the invader allele as assumed in the MF model.

**Appendix B. Model equilibria and approach to extinction**

For convenience, here and in Fig. 1, we refer to the MF model's mutual-extinction equilibrium, $(\rho_1^*, \rho_2^*) = (0, 0)$, as "phase 0." Similarly, we refer to competitive exclusion of allele 2 by allele 1, $(\rho_1^*, \rho_2^*) = (1 - \mu/\alpha_1, 0)$, as "phase 1." We refer to competitive exclusion of allele 1, $(\rho_1^*, \rho_2^*) = (0, 1 - \mu/\alpha_2)$ as "phase 2." The transition between phase 1 and mutual extinction (phase 0) is continuous in that the density of allele 1 vanishes continuously as $\alpha_1 \to \alpha_c$ from above, where the critical rate is $\alpha_c = \mu$. In particular, in the vicinity of the critical point, the equilibrium density of allele 1 increases linearly with $\alpha_1$, $\rho_1^* \approx (\alpha_1 - \alpha_c)/\mu$. The transition between phase 2 and phase 0, invader extinction, is identical by symmetry.

In the absence of mutation ($\phi = 0$), the transition between the two positive equilibria, phase 1 and phase 2, is discontinuous. The densities exhibit a "jump" of size $1 - \mu/\alpha$ crossing the line of competitive equivalence, where clonal propagation rates of the two alleles are equal: $\alpha_1 = \alpha_2 = \alpha$. For rare mutation ($0 < \phi << 1$), this transition becomes a very narrow crossover (width of order $\phi$) through which the densities change sharply, but continuously, between phases 1 and 2. Away from this crossover region (so that $|\alpha_2 - \alpha_1| >> \phi$), the densities are well approximated by those of the case where $\phi = 0$, with small corrections (again, of order $\phi$).

The main text reports that simulations of the spatial model reveal a continuous transition between either phase 1 or phase 2 and phase 0 (i.e., transition from positive equilibrium to extinction) as $\alpha_g$ is decreased. This transition occurs between (what can be termed) an active state and a single absorbing state, extinction. Therefore, it is expected to fall into the directed percolation (DP) universality class (see Hinrichsen, 2000), a technical point that leads to an ecological generality. The lattice model involves three distinct, elementary states [resident, invader, and empty sites (holes); see expression (2) in the text]. But consider the dynamics for $\alpha_2 < \alpha_c$ with $\alpha_1$ varied, where the continuous phase transition between the resident allele's persistence and extinction occurs. Since the mutation rate is small, allele 2 plays essentially no role in the dynamics. The model effectively translates to a one-allele (particles and holes) system governed by the local propagation and mortality rates of the single allele. Oborny et al. (2005) recently showed that this sort of one-type spatial model is equivalent to the contact process (Harris, 1974; Durrett and Levin, 1994b; Levin and Pacala, 1997). The latter has been investigated thoroughly using MC and renormalization-group methods of probability theory and statistical physics, and has been shown to belong to the DP universality class (Marro and Dickman, 1999; Hinrichsen, 2000), where spatial correlations between population density fluctuations can span the full system as the critical point is approached (Stanley, 1971). Consequently, MF, PA, and similar approximations based on truncating correlations, will break down (see Petermann and De Los Rios, 2004). Values we found for the critical propagation rate ($\alpha_c \approx 1.65\mu$) and the power-law exponent of the density of the extant allele near extinction ($\beta \approx 0.58$) agree very well with those of the single-species model of Oborny et al. (2005). We further verified that fluctuations of the global density rapidly increase (also in a power-law fashion) as the critical point is approached, indicating that this transition is "critical." These fluctuations imply the possibility of stochastic extinction even above the critical colonization rate for any finite system, a property underlying the ecological significance of dispersal-limited extinction (Harris, 1974; Oborny et al., 2005).

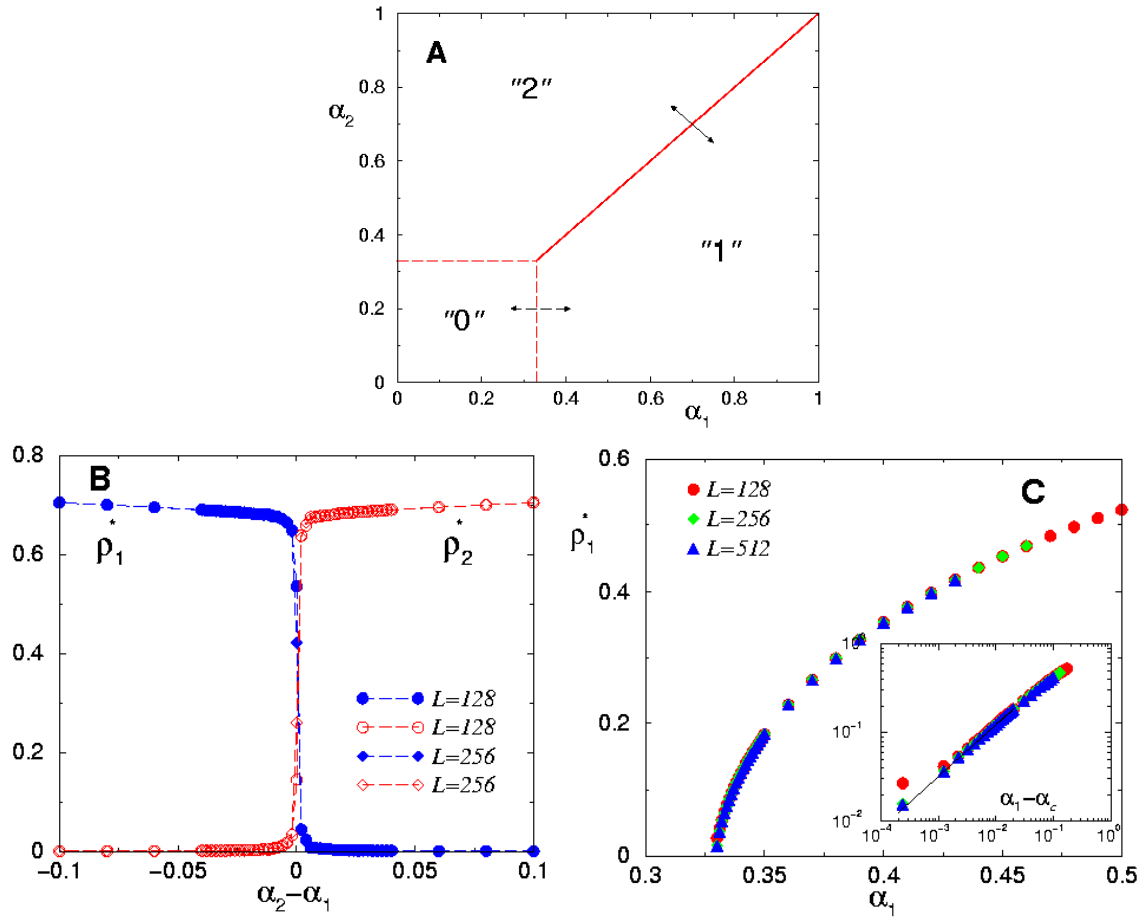

**Fig. 1** (A) Equilibrium phase diagram of the two-allele system on the $\alpha_1 - \alpha_2$ plane for $\mu = 0.20$ and $\phi = 10^{-5}$. The arrows indicate where the phase boundaries were crossed in figures (B) and (C). (B) Sharp crossover between the two active (i.e. competitively exclusive) phases, perpendicularly cutting across the $\alpha_1 = \alpha_2$ symmetry axis at $\alpha_1 = \alpha_2 = 0.70$, controlled by the difference $\alpha_2 - \alpha_1$. (C) Continuous phase transition between the active and the absorbing phases. MC simulation data is shown for various system sizes for fixed $\alpha_2 = 0.20$ with $\alpha_1$ varied. The inset indicates the power-law behavior of the density of allele 1, consistent with the DP universality class.

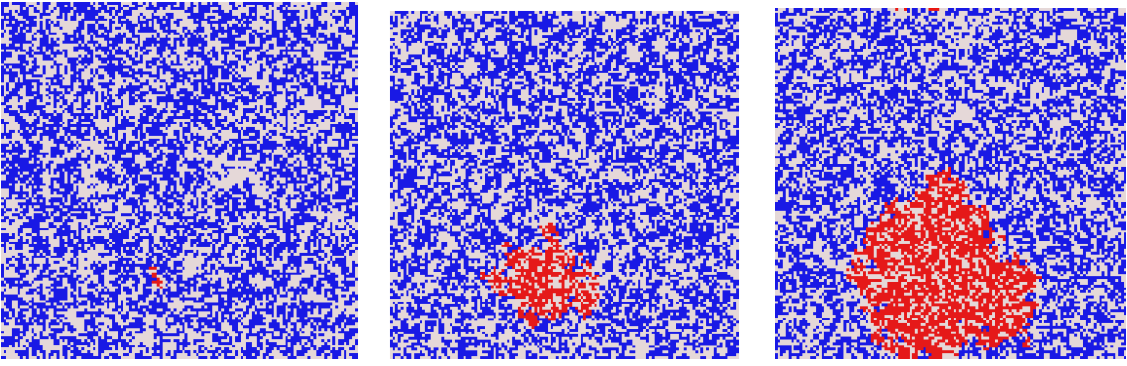

**Fig. 2** Single-cluster invasive spread. Configuration snapshots from Monte Carlo simulations for $L=128$, $\alpha_1 = 0.50$, $\alpha_2 = 0.70$, $\mu = 0.20$, and $\phi = 10^{-7}$. White represents empty sites, blue and red correspond to sites occupied by the resident and the invasive allele, respectively. Configurations are shown (from left to right) at $t = 2100$, $t = 2300$, and $t = 2500$ (in units of MCSS).

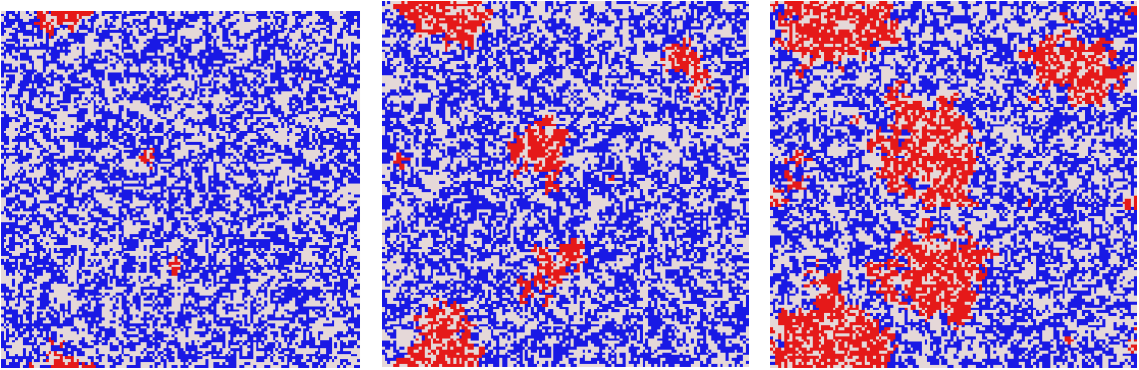

**Fig. 3** Multi-cluster invasive spread. Configuration snapshots from Monte Carlo simulations for $L=128$, $\alpha_1 = 0.50$, $\alpha_2 = 0.70$, $\mu = 0.20$, and $\phi = 10^{-5}$. White represents empty sites, blue and red correspond to sites occupied by the resident and the invasive allele, respectively. Configurations are shown (from left to right) at $t = 100$, $t = 200$, and $t = 300$ (in units of MCSS).

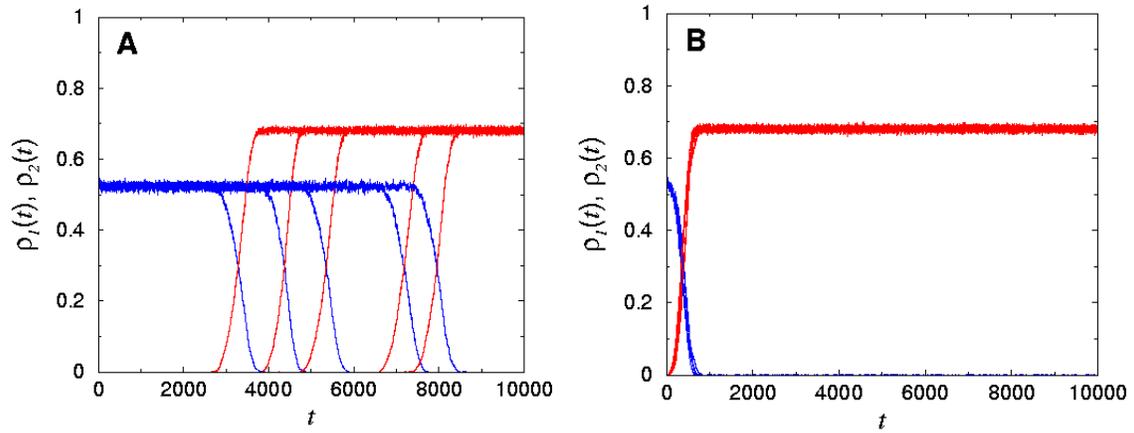

**Fig. 4** Time-dependent global densities (five independent realizations) of the two alleles in (A) single-cluster and (B) multi-cluster invasion. Matching pairs of $\rho_1(t)$ and $\rho_1(t)$ intersect near a density of 0.3. Parameters are the same as those in figure 2 and figure 3 for (A) and (B), respectively.

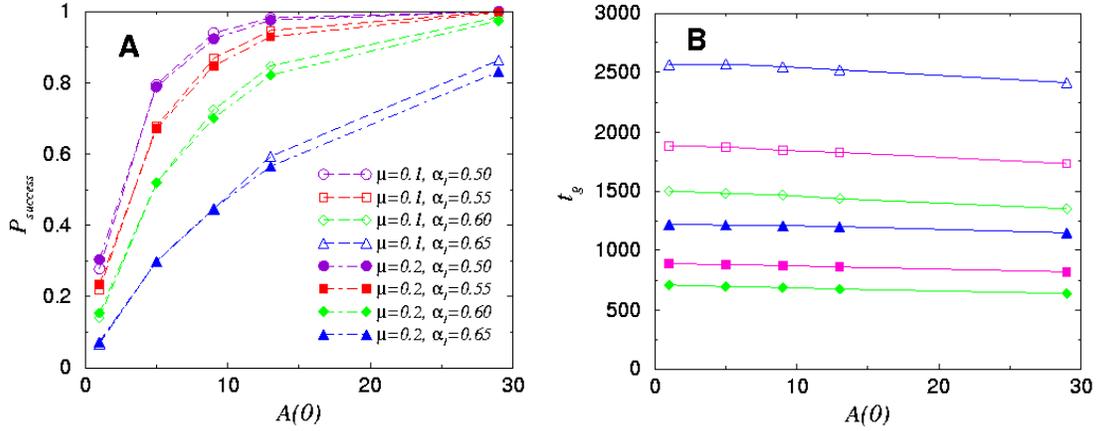

**Fig. 5** (A) Fraction of successful invasions vs. initial size of the cluster of the favored allele for two different values of the mortality rate [$\mu = 0.1$ (open symbols), $0.2$ (solid symbols)] and four different values of the propagation rate of the resident allele ($\alpha_1 = 0.5, 0.55, 0.60, 0.65$) when invader propagation rate $\alpha_2 = 0.7$. $L = 64$. Abscissa shows size (number of sites) of single invasive-allele cluster at initiation of simulation. Ordinate is the fraction of $10^4$ simulations resulting in successful advance of favored allele when further mutation has been eliminated. At critical cluster size invasion fraction is 0.5, by definition. Critical cluster size increases as difference between invader and resident propagation rates declines. Increasing the common mortality rate (0.1 to 0.2) has little effect on critical cluster size for given $\alpha_g$ values. (B) Invasion times for successful invasions. Abscissa shows size of single invasive-allele cluster at initiation of simulation. Ordinate is time (MCSS) at which resident allele's global density first falls to half its equilibrium density in absence of favored allele. $\alpha_2 = 0.7$; $L = 64$. ($\alpha_1,\mu$) combinations are (0.6, 0.1) [open diamonds], (0.625, 0.1) [open squares], (0.65, 0.1), [open triangles]; (0.6, 0.2) [solid diamonds], (0.625, 0.2) [solid squares], and (0.65, 0.2), [solid triangles]. Given the $\alpha_g$ values, increased mortality accelerates invasion dynamics and decreases time at which favored allele attains competitive superiority.

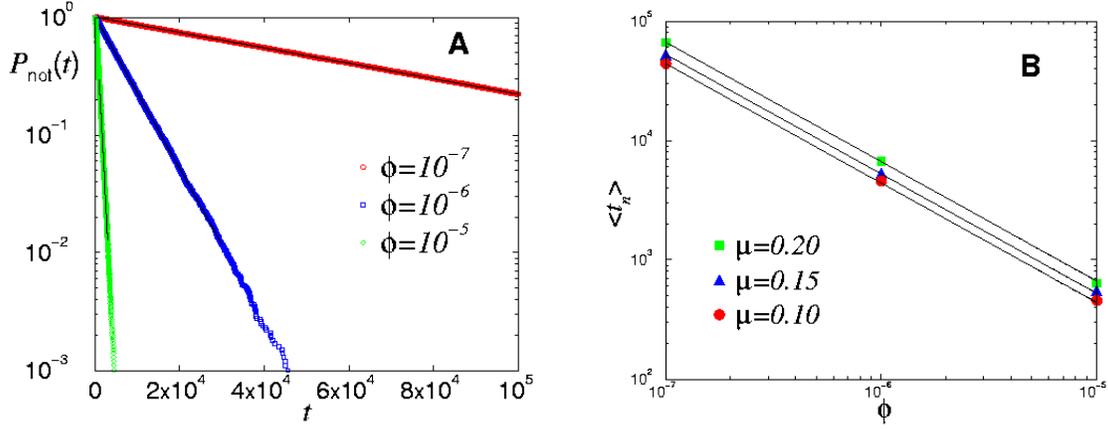

**Fig. 6** (A) Cumulative lifetime distributions in single-cluster invasion for $L = 32$, $\alpha_1 = 0.50$, $\alpha_2 = 0.70$, and $\mu = 0.20$, for three different values of $\phi$. (B) Average nucleation times [obtained as the inverse slopes of the exponentials distributions] as a function of the mutation rate $\phi$ for three different values of $\mu$. The straight solid lines correspond to $\langle t_n \rangle \sim \phi^{-1}$.

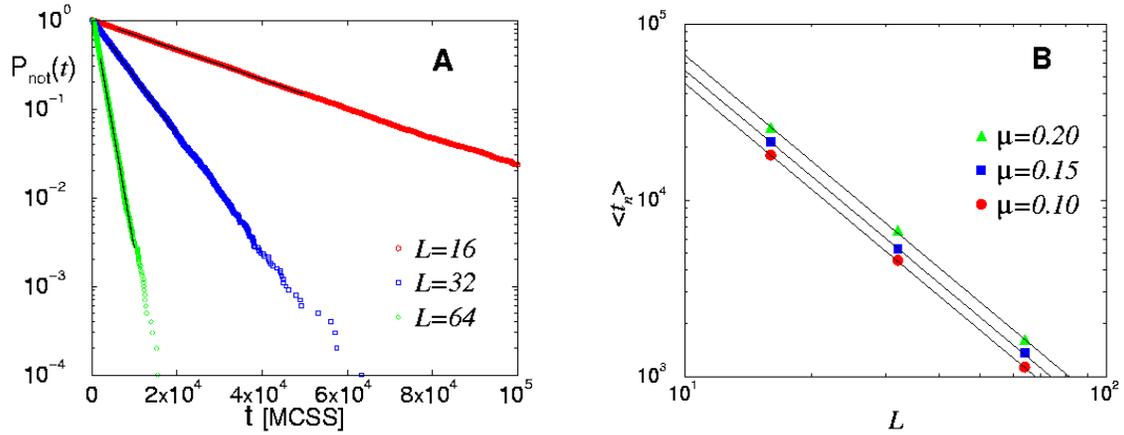

**Fig.7** (A) Cumulative lifetime distributions in single-cluster invasion for $\phi = 10^{-6}$, $\alpha_1 = 0.50$, $\alpha_2 = 0.70$, and $\mu = 0.20$, for three different values of $L$. (B) Average nucleation times [obtained as the inverse slopes of the exponentials] as a function of the linear system size $L$. The straight solid lines correspond to $\langle t_n \rangle \sim L^{-2}$.

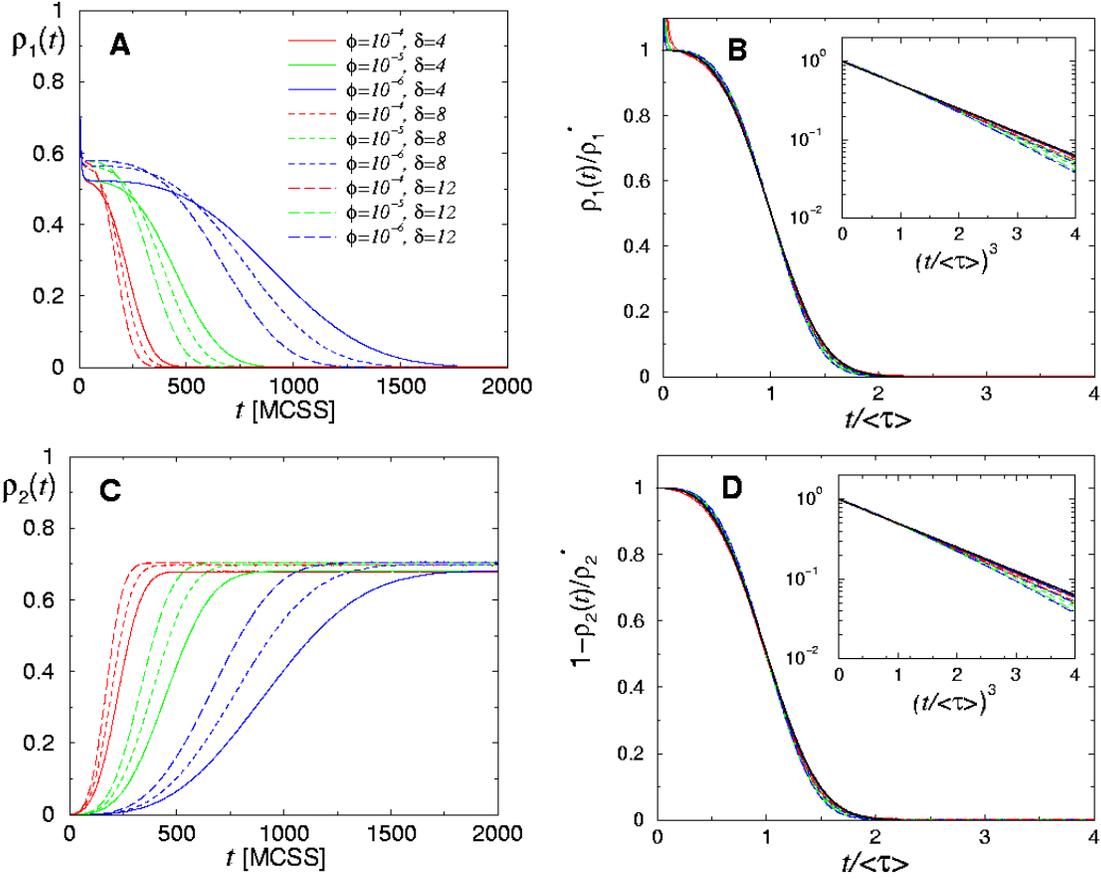

**Fig. 8** Time-dependent global densities in the multi-cluster regime for $L = 1000$, $\alpha_1 = 0.50$, $\alpha_2 = 0.70$, and $\mu = 0.20$, for various neighborhood sizes $\delta$ and mutation rates $\phi$. (A) Time-dependent global density of the resident allele. (B) Scaled time-dependent global density of the resident allele. The bold solid curve corresponds to Avrami's law Eq. (6) (hardly distinguishable from the simulation data). The inset shows the same data points, plotting the scaled density vs. $(t/\langle\tau\rangle)^3$, on log-linear scales. (C) Time-dependent global density of the invasive allele. [Symbols are the same as in (A)]. (D) Scaled time-dependent global density of the invasive allele. The bold solid curve corresponds to Avrami's law Eq. (6). The inset shows the same data points, plotting the scaled density vs. $(t/\langle\tau\rangle)^3$, on log-linear scales.

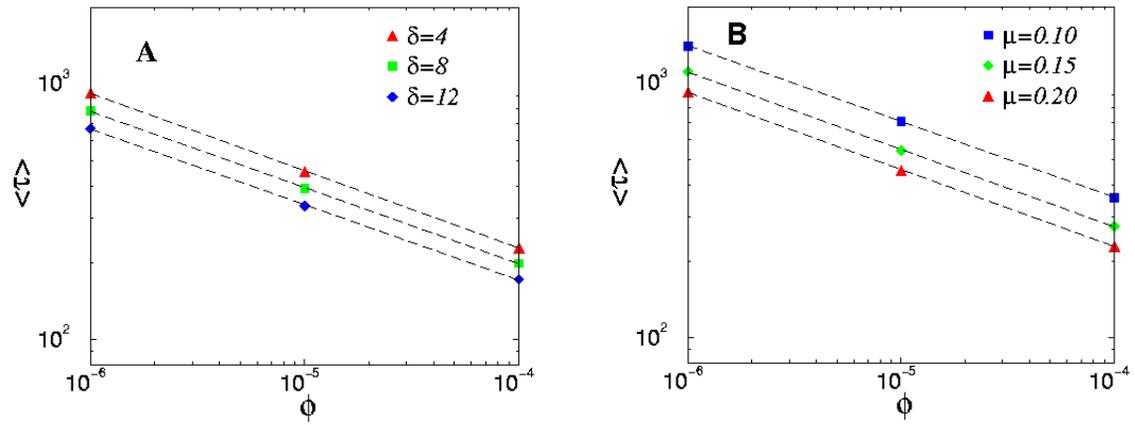

**Fig. 9** Average lifetime in the multi-cluster regime (in units of MCSS) vs. the mutation rate $\phi$ on log-log scales for $L = 1000$, $\alpha_1 = 0.50$, $\alpha_2 = 0.70$. (A) for $\mu = 0.20$ for various neighborhood sizes $\delta$. (B) for $\delta = 4$ for various values of $\mu$. The dashed lines are the fitted slopes, all approximately indicating $\langle \tau \rangle \sim \phi^{-0.3}$ [cf. Eq. (10)].

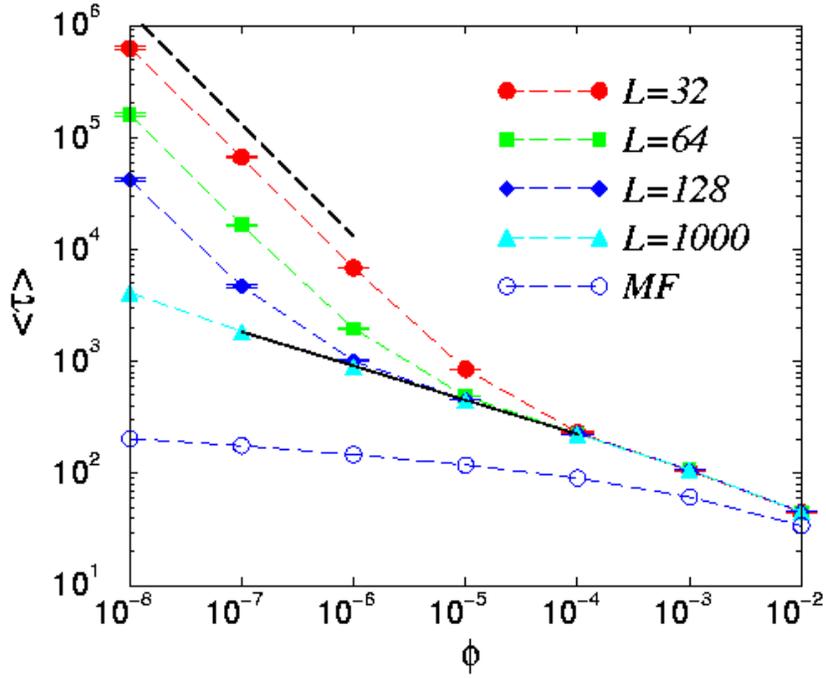

**Fig. 10** Average lifetime (in units of MCSS) vs. the mutation rate on log-log scales for $\alpha_1 = 0.50$, $\alpha_2 = 0.70$, and $\mu = 0.20$ for various system sizes $L$. The straight solid line (fitted across the $L = 1000$ system data points) is the best-fit power function indicating the $\langle \tau \rangle \sim \phi^{-0.3}$ behavior in multi-cluster regime. The straight dashed line corresponds to a slope $-1$, indicating the $\langle \tau \rangle \sim \phi^{-1}$ behavior in the single-cluster regime [cf. Eq.(9)]. For comparison, the lifetime in the mean-field approximation (MF), obtained by solving Eq. (4) numerically, is also shown.

**Table 1**
List of model symbols, definitions (numerical value or range used in the simulations, where appropriate)

| Symbol | Definition |
| --- | --- |
| $L$ | Lattice length/width ($16 \leq L \leq 1000$) |
| $n_1(\boldsymbol{x})$ | Local occupation numbers for the resident allele at site $\boldsymbol{x}$ ($n_1(\boldsymbol{x}) = 0,1$) |
| $n_2(\boldsymbol{x})$ | Local occupation numbers for the invasive allele at site $\boldsymbol{x}$ ($n_2(\boldsymbol{x}) = 0,1$) |
| $\sigma(x)$ | Set of nearest neighbors of site $\boldsymbol{x}$ |
| $\delta$ | Neighborhood size for clonal growth ($\delta = |\sigma(x)|$) |
| $\eta_g(x)$ | Density of allele $g$ on $\sigma(x)$ ($g$=1,2) |
| $\phi$ | Recurrent (forward-backward) mutation rate ($10^{-8} \leq \phi \leq 10^{-2}$) |
| $\alpha_1$ | Resident allele's clonal propagation rate |
| $\alpha_2$ | Invasive allele's clonal propagation rate |
| $\mu_1$ | Resident allele's mortality rate |
| $\mu_2$ | Invasive allele's mortality rate ($\mu_1 = \mu_2 = \mu$) |
| $\rho_1(t)$ | Resident allele's global density |
| $\rho_2(t)$ | Invasive allele's global density |
| $\rho_1^*$ | Resident's "metastable" global density |
| $\rho_2^*$ | Invader's global density at equilibrium |
| $\beta$ | Critical exponent for the global density near the continuous equilibrium phase transition associated with the dispersal-limited extinction. |
| $\langle \tau \rangle$ | Resident's "metastable" lifetime |
| $\langle t_n \rangle$ | Average waiting time for invader's nucleation |
| $t_g$ | Time for successful invader to grow to competitive dominance |
| $I$ | Nucleation rate per unit area |
| $v$ | Velocity at which cluster radius grows |
| $R_c$ | Critical radius of nucleating cluster |
| $R_d$ | Average separation of nucleating invasive clusters |